\documentstyle[twocolumn,aps,psfig]{revtex}

\def\dc{\sigma_{\rm dc}}
\def\dens{\langle n \rangle}
\def\ImL{{\rm Im}{\Lambda}}

\begin{document}
\draft
\twocolumn[\hsize\textwidth\columnwidth\hsize\csname @twocolumnfalse\endcsname 
\title{Quantum Monte Carlo Study of the
Disordered Attractive Hubbard Model}
\author{R.~T. Scalettar$^{1}$, N. Trivedi$^{2}$, 
and C. Huscroft$^{1}$}
\address{
$^{1}$Physics Department,
University of California,
Davis, CA 95616}
\address{
$^{2}$
Theoretical Physics Group,
Tata Institute for Fundamental Research,
Mumbai 400005, India}
\date{\today}
\maketitle
\begin{abstract}
We investigate the disorder-driven superconductor to insulator 
quantum phase transition (SIT) in an interacting {\em fermion}
model using determinantal quantum Monte Carlo (QMC) methods.
The disordered superconductor is modeled by an attractive Hubbard model 
with site disorder chosen randomly from a uniform distribution.
The superconducting state which exists for small disorder
is shown to evolve into an insulating phase
beyond a critical disorder.
The transition is tracked by the vanishing of (a) the superfluid stiffness, and
(b) the charge stiffness or the
delta function peak in the optical conductivity at zero frequency. 
We also show the behavior of the charge, spin, pair, and current correlations
in the presence of disorder.
Results for the temperature dependence of the dc conductivity, 
obtained by an approximate analytic continuation 
technique, are also presented both
in the metallic phase above $T_c$ and the
insulating phase.
We discuss some of the complications in extracting 
the resistance at the transition point.
\end{abstract}

\vskip2pc]
\narrowtext

\section{Introduction}

\bigskip

In a wide variety of two dimensional disordered systems\cite{REVIEW-expt},
from granular and homogeneously disordered 
Bi, Pb and Sn films\cite{GOLDMAN,DYNES,VALLES}, 
to In$_{1-x}$O$_{x}$\cite{HEBARD} and MoGe films\cite{YAZDANI}, 
high temperature superconducting films\cite{DYNES-hitc,ROSENBAUM}
and Josephson junction arrays\cite{MOOIJ}, 
a transition from a superconductor to
an insulator (SIT) can be driven by adjusting some tuning parameter such as
the film thickness, the O concentration, the magnetic field strength, 
or the charging energy.
The experimental signature of the transition is that
the behavior of the sheet resistance $R_{\Box}(T)$ as a function 
of temperature $T$ 
is different in the two phases. 
At low disorder or magnetic field, the system is superconducting for $T<T_c$.
The transition temperature $T_c$
decreases with increasing disorder or magnetic field
and above $T_c$ the system is metallic with $d R_{\Box}/dT >0$. 
Beyond a critical disorder or magnetic field, on the other hand,
the system becomes insulating with $d R_{\Box}/dT<0$.

Motivated by these experiments, one of the important open theoretical 
questions is to study particular 
microscopic models to see whether or not they show a SIT
as a function of some tuning parameter such as the degree of disorder
and, if so, characterize the transition.

Anderson\cite{ANDERSON} proposed that the superconducting 
transition temperature $T_c$ 
and the thermodynamic properties should be {\em unaffected} by disorder
since Cooper pairs can
be formed by pairing the time-reversed exact eigenstates of the
noninteracting disordered problem. 
This is only valid for small disorder in the regime
$k_F \ell \gg 1$, where $k_F$ is the Fermi momentum and $\ell$ is the
elastic mean free path. 
Ma and Lee\cite{MA-LEE} developed a mean field theory 
in which they assumed that
the order parameter was uniform throughout the system.
As a consequence, the superfluid density remained large even for fairly high
disorder and was found to persist essentially all 
the way to the site localized limit.

One might therefore ask whether a disorder--driven SIT can occur at all. 
It is important to note that both the Anderson and Ma--Lee arguments
make specific assumptions concerning the effect of randomness,
and hence may not be compelling in all cases.
In order to understand why a SIT might be possible,
consider the two generic mechanisms for the
destruction of superconductivity.  First, the magnitude of the pairing
gap can be driven to zero.  Second, phase coherence between the pairs
in different parts of the sample may be lost.  Clearly there is
an interplay between fluctuations in the pair amplitude and phase.
For example, the phase can change at a smaller energy cost in regions
where the amplitude is lower\cite{NT-bound}. 
It is possible that the pair amplitude is
driven to zero at the same point where phase coherence is lost, but it
is also possible that the two phenomena occur separately.

Fisher and collaborators\cite{FISHER-qpt}
were the first to describe a scenario in
which phase fluctuations caused a SIT while the pair amplitude
remained finite.  They conjectured that the SIT might be
in the same universality class
as the superfluid-insulator transition for bosons.
They argued that since near the transition the size of the Cooper pair
is much smaller than the diverging correlation length, it is possible to
describe it as a bose field.
Of course, the charge carriers of the experimental systems are 
fermionic in nature, so it is useful to study Hamiltonians
which do not begin immediately with bosonic degrees of freedom.
Perturbative methods to study the SIT in fermionic models
have not been successful in describing the transition 
region\cite{Finkelstein,Belitz}, which is
not surprising since the transition
occurs in a region of high disorder 
in an interacting system.

While this approach has led to a number of very interesting results,
especially for the value of the conductivity at the 
transition\cite{CHA,SORENSEN,RUNGE,BATROUNI},
it is important to test the validity of the phase-only models
by developing methods which also treat amplitude fluctuations.
In order to better describe the behavior of a superconductor at high disorder,
Ghosal and collaborators\cite{GHOSAL} have included the fluctuations of
the superconducting order parameter by solving 
the ``Bogoliubov-de Gennes" mean field equations self-consistently.
They have found that the probability distribution of the 
local pairing amplitude 
develops a broad distribution with significant weight near zero with 
increasing disorder.
Surprisingly, the density of states continues to show a finite spectral
gap, as also seen by Quantum Monte Carlo (QMC) and maximum entropy 
techniques\cite{HUSCROFT-MAXENT}, shown to arise from 
the break up of the system into superconducting islands separated
by regions with very small pairing amplitude.
These disorder-induced fluctuations in the order parameter amplitude 
have a marked effect in
suppressing the superfluid density at higher disorder
but by themselves are not sufficient to drive the system non-superconducting.
It is necessary to include phase fluctuations distributed inhomogeneously
riding on top of the highly inhomogeneous amplitude fluctuations
to get a SIT.

In this paper we describe
the first QMC study of a fermion model of superconductivity (the 
attractive Hubbard Hamiltonian with random site energies) 
which gives a SIT at a critical disorder strength\cite{NT-sit}.
The attractive Hubbard Hamiltonian which we study
is a simple model of a disordered SC that allows us to explore the 
qualitative issues arising
from the interplay of superconductivity and localization.
While such a model does not address questions concerning the 
microscopic origin of the pairing, since the attraction is 
put in {\it a priori}, one can nevertheless examine
questions such as the competition between superconductivity and
charge density wave formation\cite{HUSCROFT}, 
the behavior of superconducting correlations
above the superconducting transition temperature 
\cite{NT-pseudogap1,NT-pseudogap2,NT-pseudogap3,RANDERIA-varenna,NT-sns97},
and the interpolation
between weak coupling BCS and strong coupling bosonic regimes
of pair formation\cite{RANDERIA-crossover}.

\section{Organization of the Paper}

This paper is organized as follows:
In section III we introduce the attractive Hubbard model
and briefly review the
physics of the clean attractive Hubbard model.
In Section IV we describe the QMC simulation
technique.
In section V we first discuss our results for the chemical potential,
in order to demonstrate that we are in the degenerate Fermi regime of the 
model. We then describe the effect of disorder on the local and longer range
density-density and pairing correlations.
The pairing correlations are found to be much more robust compared to the
density correlations away from half filling.
We also show the behavior of the superconducting order parameter
which decreases rapidly with increasing disorder and vanishes beyond
a critical disorder.
In section VI we present a detailed discussion of the 
longitudinal and transverse current-current correlation functions. 
The longitudinal response obeys 
the f-sum rule and equals the absolute
value of half the lattice kinetic energy $K_x$
which we verify in our simulations.
The transverse response on the other hand, deviates from $K_x$
and this deviation is a measure of the 
superfluid stiffness of the system. We present results showing the suppression
of the superfluid stiffness with disorder and its ultimate destruction 
beyond a critical disorder.
In section VII we discuss the behavior of the frequency dependent
current-current correlation function and the extraction of the charge
stiffness or the strength of the delta-function peak in the optical
conductivity. Our results show that in the superconducting phase, the 
superfluid stiffness and the charge stiffness are roughly equal in 
magnitude for all disorder strengths.
In section VIII we discuss an approximate method to extract the 
temperature dependence of the dc resistivity and show its behavior in the 
metallic phase above $T_c$ for low disorder as well as in the insulating phase
for higher disorder.
The resistivity at the transition is extracted by two methods--
(i) At the critical disorder, the charge stiffness vanishes with 
frequency with a slope proportional to the resistivity;
and (ii) from the crossing of the resistivity vs disorder
curves at various temperatures.
We also discuss the complications of obtaining the
resistivity near a quantum critical point.
We present our conclusions in section IX and end with 
some of the outstanding questions in the area of SIT in section X.
In previous papers\cite{NT-sit,NT-dae,RTS-mrs} we have presented
a short discussion of some of these issues.
The purpose of the present manuscript is
to provide the details behind that work, as well
as to present a number of new results including a more
complete discussion of both the physics and the numerics.

\section{Model}

The Hamiltonian we study is defined by,
\begin{eqnarray}\label{ham}
H &=& -t\sum_{\langle {\bf ij} \rangle \sigma} 
( c_{{\bf i} \sigma}^{\dagger}c_{{\bf j} \sigma}
+ c_{{\bf j} \sigma}^{\dagger}c_{{\bf i} \sigma} ) \nonumber\\
& &- \sum_{{\bf i} \sigma} (\mu - V_{{\bf i}}) n_{{\bf i} \sigma}
- |U| \sum_{{\bf i}} (n_{{\bf i} \uparrow}-\frac12) 
                    (n_{{\bf i} \downarrow}-\frac12)\  \cdot
\label {eq:hamil}
\end{eqnarray}
Here the lattice sum $\langle {\bf ij} \rangle$ is over
nearest neighbor sites on a two dimensional square lattice,
$c_{{\bf i}\sigma}$ is a fermion destruction operator at site ${\bf i}$
with spin $\sigma$, $n_{{\bf i}\sigma}=
c_{{\bf i}\sigma}^{\dagger}c_{{\bf i}\sigma}$,
and the chemical potential $\mu$ fixes the average density $\dens$.  
The site energies $V_{{\bf i}}$ are independent random variables with a 
uniform distribution over $[-V/2,V/2]$.  
The interaction has been written in particle-hole
symmetric form so that $\mu=0$ corresponds to $\dens=1$ at
all $U$ and $T$ when $V=0$.  
We set $t = 1$ and measure all energies in units of $t$.

In real materials, disorder plays a complicated role in
the Hamiltonian, both affecting the screening of the electron-electron
interaction as well as the phonons and hence the electron-phonon interaction.
Our Hamiltonian does not include these effects.


Some of the physics of the clean attractive Hubbard model may be summarized
as follows\cite{RTS-MOREO,MICNAS}:  At half-filling ($\mu=0$), the model has no
long range correlations at any finite temperature, and 
at $T=0$ is in a state with combined 
charge density wave (cdw) and superconducting order\cite{PH}.
When $\mu \neq 0$ the system has a finite temperature Kosterlitz-Thouless
transition to a state with superconducting order.
The transition temperature $T_c$ depends strongly on the filling near
$\dens=1$.  $T_c$ shows a non-monotonic dependence on 
coupling\cite{RANDERIA-crossover}
similar to the repulsive Hubbard model where the
Neel temperature first increases with $U$ but then goes down
as $T_{N} \propto J = 4t^{2}/U$ at strong coupling. 
Numerical estimates\cite{MOREO-tc,DENTENEER} 
of $T_c$ are still a matter of considerable debate and
at $\dens=0.875$ vary from $0.3t$ to $0.03t$.

\section{Quantum Monte Carlo Simulation}

Our simulation uses the standard ``determinant'' QMC
algorithm\cite{BSS,SCALAPINO-review}, along with its various 
refinements\cite{SORELLA,WHITE}.
The partition function $Z={\rm Tr} [e^{-\beta H}]$ is written as a
path integral by discretizing the imaginary time dimension
$\beta=1/T$ into $N_\tau$ time slices as 
\begin{eqnarray}
e^{-\beta H} &=& \left(e^{-\Delta\tau H}\right)
^{N_\tau}\nonumber\\
& \approx & \left(e^{-\Delta\tau H_1} e^{-\Delta\tau H_U}\right)^{N_\tau}
\label {eq:trotter}
\end{eqnarray}    
where $\beta=N_\tau \Delta\tau$. In Eq.~\ref{eq:trotter},
$H_1$ is the sum of the two single particle terms in 
Eq.~\ref{eq:hamil} and $H_U$ is the interaction term. 
A systematic Trotter error is introduced in Eq.~\ref{eq:trotter}
because of the non-commutativity
of the operators $H_1$ and $H_U$.
This Trotter error, however, can be dealt with, either by making 
$\Delta \tau$ sufficiently small so that errors in observables are of 
the same order as statistical fluctuations from the sampling, or, if 
greater accuracy is needed, by extrapolating to $\Delta \tau = 0$.
The exponential of the interaction term is decoupled using a 
Hubbard-Stratonovich (HS) transformation by introducing a 
discrete field\cite{HIRSCH}
$S_{i\tau}=\pm 1$ at each point in the space-time lattice, 
\begin{eqnarray}
& &\exp[+\Delta\tau\mid U\mid (n_{i\uparrow}-1/2) (n_{i\downarrow}-1/2) ]
=\nonumber \\
& & \frac12 \exp\left\lbrace{-\Delta\tau{\mid U\mid}\over 4}\right\rbrace 
\sum_{{S_{i\tau}=\pm 1}}
\exp[\Delta \tau \lambda S_{i\tau}(n_{i\uparrow}+n_{i\downarrow}-1)]
\label{eq:hs}
\end{eqnarray}
where,
\begin{equation}
\cosh(\Delta\tau\lambda)=\exp(\Delta\tau
\mid U \mid/2),
\label{eq:constraint}
\end{equation}
is satisfied by {\em real} $\lambda$.
Thus the original functional integral over Grassman variables,
which involved traces containing {\em quartic} operators is reduced to 
a quadratic problem in the fermion operators {\em but at the cost of performing
a sum over all configurations of the HS fields on the
discretized space-time lattice}.
The partition function in the grand canonical ensemble is
\begin{eqnarray}
&Z&={\rm Tr} \exp\left\lbrace{-\beta H}\right\rbrace \nonumber\\
 &=&\sum_{\lbrace S\rbrace} {\rm Tr}
\prod_{\tau,\sigma} 
\exp{\left [ -\Delta\tau\sum_{i,j} c_{i\sigma}^\dagger
h_{\lbrace S\rbrace} (\tau,\sigma) c_{j\sigma}
\right]}\  \cdot\nonumber \\
& & \           
\label{eq:hs2}
\end{eqnarray}
Here $h_{\lbrace S\rbrace} (\tau,\sigma)$ is a one-body Hamiltonian 
for the motion of an electron in a given configuration of the H-S fields.
Note in Eq.~\ref{eq:hs} both the up and 
down electrons couple to the HS field with the {\em same} sign.

Now the resulting trace over quadratic forms in the fermion operators 
in Eq.~\ref{eq:hs2} is performed and gives
\begin{equation}
Z=\sum_{\lbrace S\rbrace} \det M_\uparrow (\lbrace S\rbrace)
\det M_\downarrow (\lbrace S\rbrace)
\label{eq:det}
\end{equation}
with
\begin{equation}
M_\sigma(\lbrace S\rbrace ) = \left[ I + \prod_\tau
e^{-\Delta \tau h_{\lbrace S\rbrace}(\tau,\sigma)} \right]\  \cdot
\label{eq:m}
\end{equation}
Thus the interacting problem is equivalent to solving a non-interacting problem 
for a given HS field configuration $\lbrace S_{i\tau}\rbrace$ and then 
summing over all possible configurations.
The sum over the HS fields on the space-time lattice
is efficiently done using Monte Carlo techniques
which generate the configurations,
treating the product of the determinants as a probability. 
Note that in general for a fermion problem, since the sign 
of the determinants may be negative, the product is not necessarily non-
negative and it cannot be treated as a probability. This is the origin of the
`sign-problem' for typical fermion problems. However, for the Hamiltonian 
in Eq.~\ref{eq:hamil}, 
since it is possible to couple the HS field to the charge 
$n_{i\uparrow} + n_{i\downarrow}$ 
and satisfy
Eq.~\ref{eq:constraint} with real $\lambda$,
the two determinants in Eq.~\ref{eq:det} are {\em identical}, 
and hence the integrand is non-negative-- 
thus there is no sign problem\cite{HIRSCH} in 
attractive Hubbard model simulations {\em at any filling}.

In the determinant QMC approach,
finite temperature expectation values of combinations of 
fermion operators with arbitrary space and imaginary time arguments
can be easily evaluated. More precisely, if all the operators are at 
the same imaginary time, the observables can be expressed
in terms of matrix elements of the inverse of the matrices whose determinants
give the Boltzmann weight. These matrix elements are needed to
update the HS field, and are therefore available
``free of charge'' for the measurements.  
If the operators whose expectation values are to be measured
have different imaginary time arguments,
some extra calculations are involved to obtain the
non--equal time Green's functions. However this can be
done in a straightforward manner\cite{BSS,SORELLA,WHITE}.

\section{Equal Time Correlations}

\subsection{Chemical Potential}

The location of the chemical potential relative to the bottom of the band
gives information about the degeneracy of the system.
In the simulations presented in this paper the filling is chosen to be
$\dens=0.875$ close to the point where $T_c$ is expected to be maximal for
$U=-4t$\cite{MOREO-tc}.
For a given value of the parameters--interaction strength $U$, disorder
strength $V$ and temperature $T$--the chemical potential $\mu$ is tuned 
so that upon disorder averaging
the density $\langle n \rangle \sim 0.875$. 
We comment that an alternative approach is to tune the chemical potential
for each disorder realization separately so that each has the same
desired filling.  This is likely to result in reduced 
fluctuations\cite{PAZMANDI},
but is considerably more time consuming numerically.
Some such approach, however, appears essential for analytic continuation
calculations\cite{HUSCROFT-MAXENT}.

\begin{figure}
\vskip-02mm
\hspace*{0mm}
\psfig{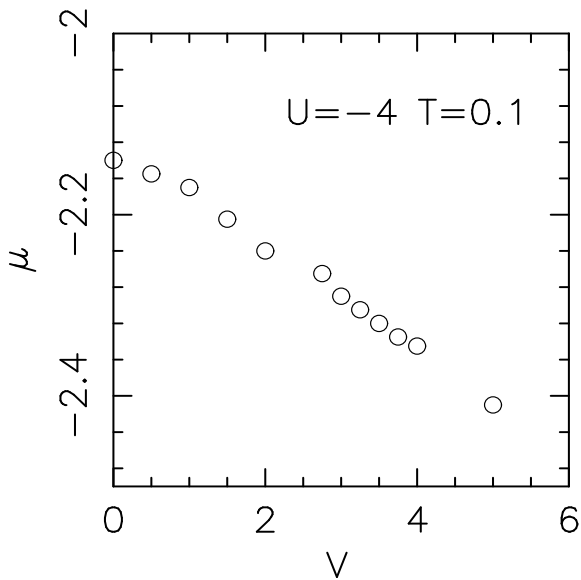}  
\vskip2mm
\caption{
The chemical potential $\mu$ shows a roughly linear decrease with 
disorder $V$. 
Since $\mu(T,|U|,V) + 4t + $ $\langle n \rangle |U|/2 \sim 3.5 t \gg T $
the system is in a highly degenerate regime
and away from the preformed bosonic regime.
}
\label{fig:mu}
\end{figure} 
The dependence
of $\mu$ on $V$ is roughly linear and is shown in Fig.~\ref{fig:mu}
for $U=-4t$.
Since $\mu$, measured from the bottom of the band
and taking into account the Hartree shift, is larger
than the temperature, 
$\mu(T,|U|,V) + 4t + \langle n \rangle |U|/2 > T $
the system is degenerate and far from the regime where
there are preformed bosons. Note, we have assumed that the bottom of the 
band is at $-4t$, which is the case in the clean system but should
be renormalized by the random potential in the disordered system. 

\subsection{Density-density correlations}

\begin{figure}
\vskip-02mm
\hspace*{-5mm}
\psfig{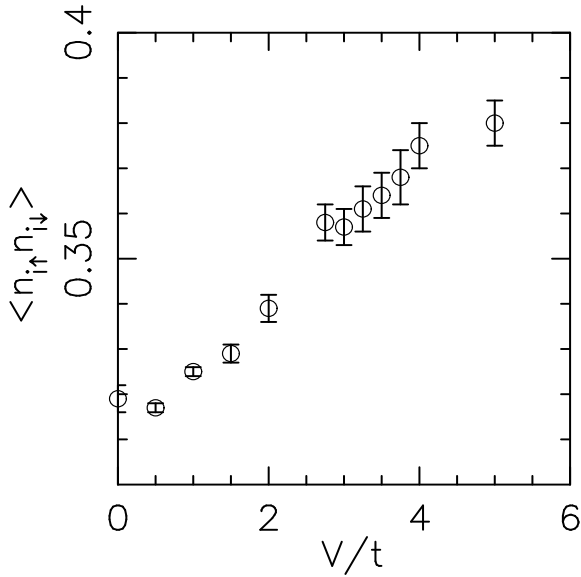}  
\vskip2mm
\caption{
The increase in double occupancy 
$\langle {\bf n_{i\uparrow}} {\bf n_{i\downarrow}}
\rangle$ as a function of increasing disorder $V/t$.
}
\label{fig:nnud0}
\end{figure}

In Fig.~\ref{fig:nnud0} we show the double occupancy 
$\langle {\bf n_{i\uparrow}} {\bf n_{i\downarrow}}
\rangle$ which is found to increase from 0.32 at $V=0$ to 0.38 at $V=5$.
This increase is a consequence of the fact that
in the attractive model, random site energies and interactions
both act to promote double occupancy, in contrast to the
repulsive model where they compete.

In Figs.~\ref{fig:nn}(a) and (b) we 
also show the spatial variation of the density-
density correlation function 
\begin{equation}
C_{\sigma,\sigma^\prime}({\bf l})=
\langle {\bf n}_{{\bf i}\sigma}{\bf n}_{{\bf i}+{\bf l},\sigma^\prime}\rangle
-\langle {\bf n}_{{\bf i}\sigma}\rangle 
\langle{\bf n}_{{\bf i}+{\bf l},\sigma^\prime}\rangle\  \cdot
\label{eq:nn}
\end{equation}
At half filling,
$C({\bf l})$ is rapidly suppressed by disorder\cite{HUSCROFT}; via finite size
scaling it is seen that even as little disorder as $V=0.25t$ is capable of
destroying the charge density wave ordering 
and in an $8\times 8$ system $C({\bf l})$ is
definitely suppressed by $V=1t$.
Away from half filling even for the clean system $C({\bf l})$ is small
and thereafter disorder does not 
have any further effect.

\subsection{Pair Correlations}

An important characteristic of the superconducting state
is that the equal time s--wave pair correlation function
$P_s$ defined by,
\begin{eqnarray}
P_s ({\bf l}) &=& \langle \,\, \Delta_{{\bf i}}^{ \,}
\Delta_{{\bf i+l}}^{\dagger} \,\, \rangle,
\nonumber\\
\Delta_{{\bf i}}^{\dagger}  &=& c_{{\bf i} \uparrow }^{\dagger}
c_{{\bf i} \downarrow }^{\dagger},
\label {eq:pair}
\end{eqnarray}   
has a finite value at large separations 
$P_s({\bf l} = (L/2,L/2))= \Delta_{OP}^2$, 
where $\Delta_{OP}$ is the ``order parameter''
on a lattice of finite size $L$.

\begin{figure}
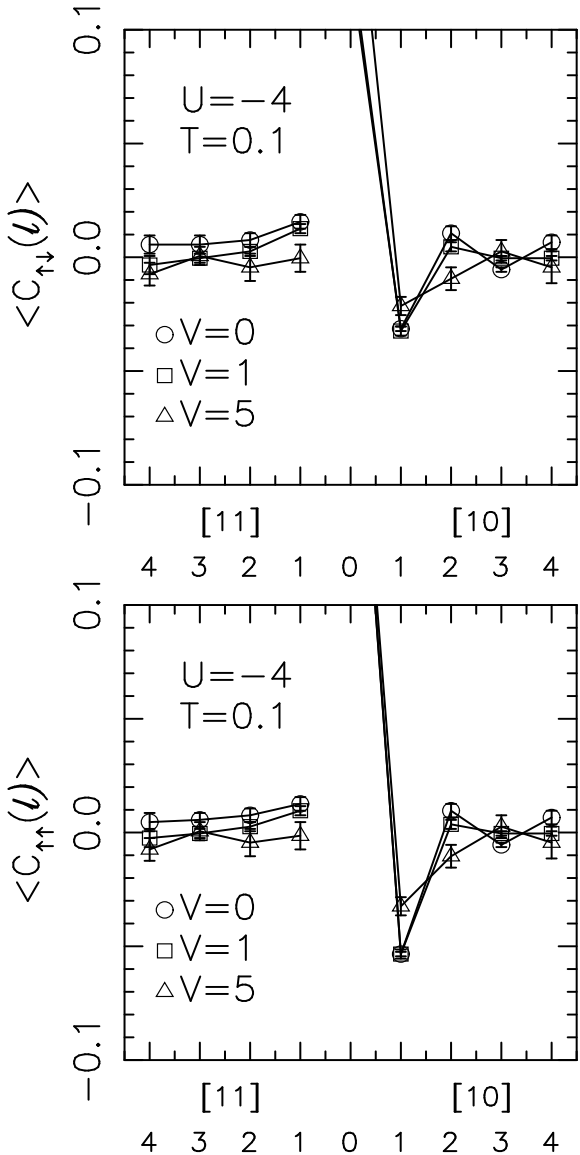

\vskip-00mm
\hspace*{-1mm}
\psfig{file=nnud.ps,height=3.0in,width=3.0in,angle=-90}  
\vskip-00mm
\hspace*{-1mm}
\psfig{file=nnuu.ps,height=3.0in,width=3.0in,angle=-90}  
\vskip+1mm
\caption{
The density correlation function $C_{\sigma,\sigma^\prime}({\bf l})$ from 
Eq.~\ref{eq:nn} 
for ${\bf l}$ along [10]
and [11] directions for 
(a) $(\sigma,\sigma^\prime)=(\uparrow, \downarrow)$
and (b) $(\sigma,\sigma^\prime)=(\uparrow, \uparrow)$,
showing rapid suppression with increasing disorder.
}
\label{fig:nn}
\end{figure} 

\begin{figure}
\vskip-00mm
\hspace*{-2mm}
\psfig{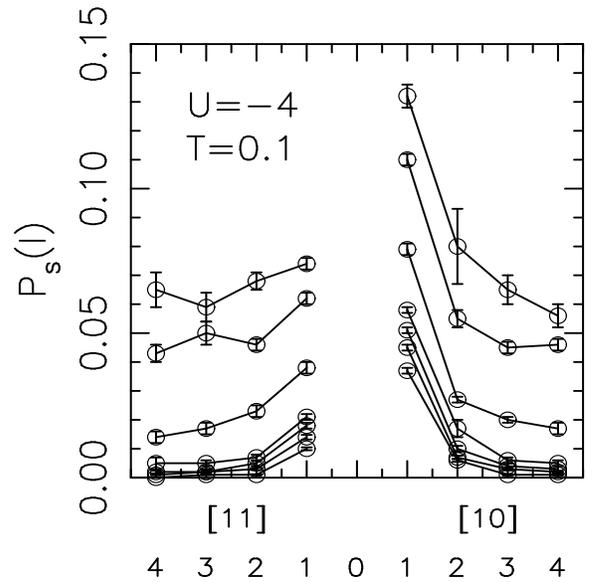}  
\vskip+2mm
\caption{
The pair correlation function defined in Eq.~\ref{eq:pair} is
shown as a function of ${\bf l}$, the relative separation of the two sites
along [10] and [11] directions for varying disorder strengths
$V=$0, 1.0, 2.0, 3.0, 3.5, 4.0, and 5.0.
The value at ${\bf l}=0$ is given by Eq.~\ref{eq:pair0} but is not shown
as it is off-scale.
Note the relative robustness of the pairing correlations 
compared to the density correlations in Fig.~\ref{fig:nn} in the presence
of disorder.
}
\label{fig:pair}
\end{figure}

In Fig.~\ref{fig:pair} we show the behavior of $P_s$ 
at a temperature $T=0.1t$ for varying
degrees of disorder. 
This temperature is sufficiently low that for the clean system the
correlation length has exceeded the linear lattice size 
and the system is effectively in the ground state.
For the clean system, or weak disorder, the correlation function
approaches a constant at large distances, implying a SC state with
long range order. For strong disorder, the correlation function
vanishes at large distances indicating the absence of an order parameter.
It is evident by comparison with Fig.~\ref{fig:nn} that pairing 
correlations
are much more robust than density--density correlations
for the same degree of disorder, as in the half filled case
\cite{HUSCROFT}.

\begin{figure}
\vskip0mm
\hspace{0mm}
\psfig{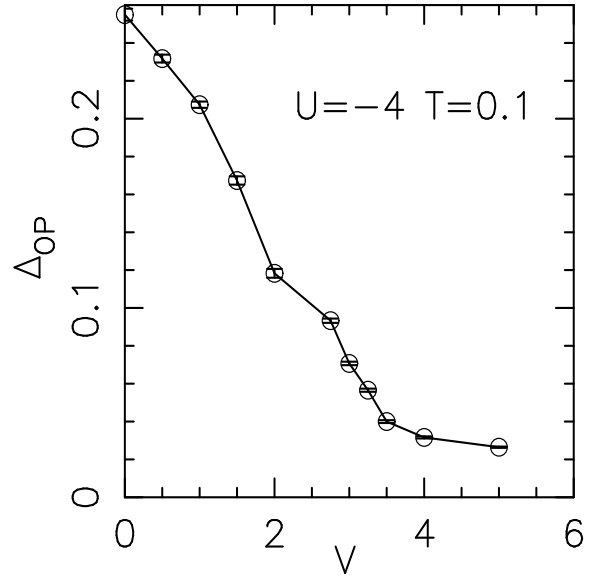}  
\vskip+2mm
\caption{
Suppression of the superconducting ``order parameter''
$\Delta_{OP}$ on an 8x8 lattice with increasing disorder.  While $\Delta_{OP}$
does not vanish at large $V$ due to finite size effects, 
a scaling analysis of the pair structure factor
indicates that in the thermodynamic limit
$\Delta_{OP}$ vanishes around a critical
disorder $V_c\sim 3.5t$.
}
\label{fig:pairmax}
\end{figure} 

Fig.~\ref{fig:pairmax}
shows the order parameter $\Delta_{OP}$
as a function of disorder which is strongly suppressed by disorder
and vanishes beyond a critical disorder strength $V_c\sim 3.5t$.

The value of the pairing correlation function at zero separation
is related to the occupancy and double occupancy,
\begin{eqnarray}
P_s({\bf 0}) &=& \langle\Delta_{\bf i} \Delta_{\bf i}^\dagger\rangle
\nonumber\\
&=& 1-\langle n\rangle + \langle n_{i\uparrow}n_{i\downarrow}\rangle\ .
\label{eq:pair0}
\end{eqnarray}
Whereas $P_s({\bf l})$ is reduced by disorder for ${\bf l}$ nonzero,
$P_s({\bf 0})$ is increased, since the density $\langle n \rangle$ is
fixed and the double occupancy rate $\langle n_{i\uparrow}n_{i\downarrow}
\rangle$ is increased (Fig.~2).


\begin{figure}
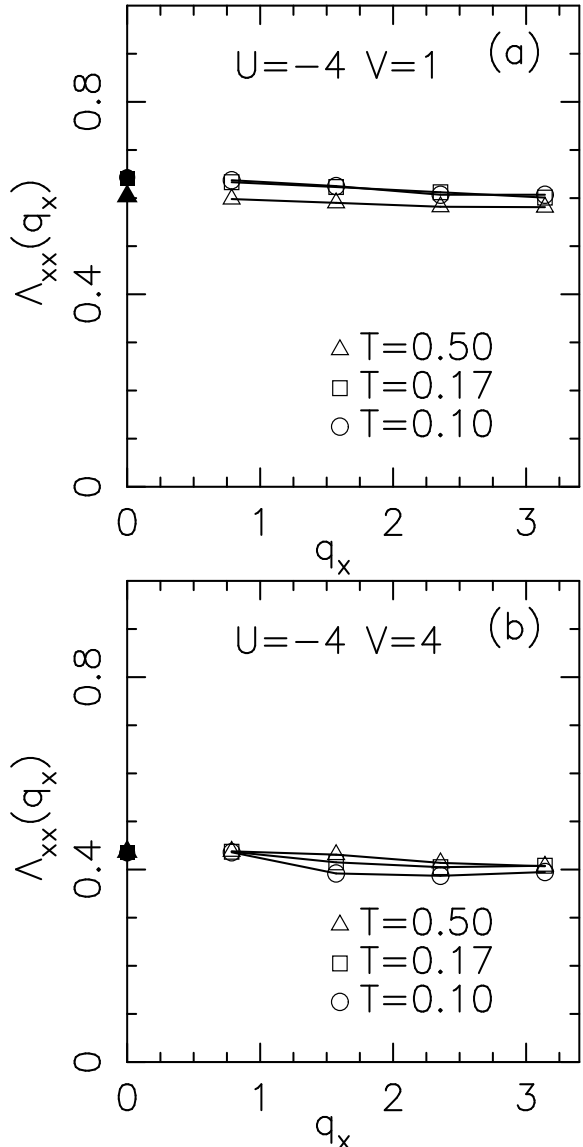

\vskip0mm
\hspace*{0mm}
\psfig{file=lamLV1.ps,height=3.0in,width=3.0in,angle=-90}  
\vskip0mm
\hspace*{0mm}
\psfig{file=lamLV4.ps,height=3.0in,width=3.0in,angle=-90}  
\vskip0mm
\caption{
The longitudinal current-current correlation function $\Lambda_{xx}(q_x)$
defined 
in Eq.~\ref{eq:long}
as a function of $q_{x}$
at $T=0.5t$ (open triangles), $0.17t$ (open squares), and $0.1t$
(open circles).
The corresponding filled points at $q_x=0$ are the 
magnitude of the kinetic energy $K_x$ along $x$ at those temperatures.
In (a) $V=1t$ and in (b) $V=4t$.
In all cases $\Lambda^{\rm L}=\Lambda_{xx}(q_{x} \rightarrow 0)$ 
approaches $K_x$ as required by gauge invariance.
}
\label{fig:lamL}
\end{figure} 

The equal time pair and density correlations already give considerable insight 
into the effect of disorder on superconductivity. The long range pairing
order in the ground state is suppressed to zero 
for disorder $V\sim 4t$, when $U=-4t$. Off half-filling, the charge
correlations are small and little affected by randomness, though disorder
does cause an enhancement of the double occupancy rate. However, considerably
more information can be obtained by looking also at various imaginary time 
dependent quantities such as the current-current correlation function.
\section{Current-Current Correlation Function}

As known for some time\cite{BAYM}, and also described recently
in the context of quantum simulations\cite{SWZ}, various limits of
the current--current correlation function give information
about the charge and superfluid stiffness, and 
gauge invariance, and in principle can be used 
to distinguish insulators, metals, and superconductors.
The current-current correlation function $\Lambda_{xx}(\bf {l},\tau)$
is defined by
\begin{eqnarray}
&{\Lambda}&_{xx}({\bf l}, \tau)
= \langle j_{x} ({\bf l}, \tau) j_{x} (0, 0) \rangle
\nonumber\\
&j&_{x}({\bf l} \, \tau) = e^{H \tau}
\left[it \sum_\sigma
(c_{{\bf l}+\hat x,\sigma}^{\dagger}c_{{\bf l},\sigma}^{\ } -
c_{{\bf l},\sigma}^{\dagger} c_{{\bf l}+\hat x,\sigma}^{\ } )
\right] e^{-H \tau}     
\label {eq:lam}
\end{eqnarray}
Upon Fourier transforming in space and imaginary time we get
$\Lambda_{xx}({\bf q},\omega_n) = \sum_{\bf l} \int_0^\beta d \tau
e^{i {\bf q}\cdot{\bf l} }
e^{-i \omega_n \tau}
\Lambda_{xx}({\bf l},\tau)$,
where $\omega_n = 2n\pi/\beta$.   
                                            
\subsection{Longitudinal response}

The longitudinal part
of $\Lambda_{xx}$ defined in Eq.~\ref{eq:lam} must satisfy the f-sum rule,
\begin{eqnarray}
\Lambda^{{\rm L}} & \equiv & 
{\rm lim}_{q_x \rightarrow 0} \hskip0.1in   
{\Lambda}_{xx} (q_{x},q_{y}=0,\omega_n = 0) \nonumber\\
\Lambda^{\rm L} &=&
K_{x}
\label {eq:long}
\end{eqnarray}
as a consequence of gauge invariance\cite{BAYM,SWZ}.
Here $K_{x} = \langle t \sum_\sigma
(c_{{\bf l}+\hat x,\sigma}^{\dagger}c_{{\bf l},\sigma}^{\ } +
c_{{\bf l},\sigma}^{\dagger} c_{{\bf l}+\hat x,\sigma}^{\ } ) \rangle$ is
the magnitude of the kinetic energy in the $x$ direction.

Fig.~\ref{fig:lamL} shows $\Lambda_{xx}(q_x)$ as a function of $q_x$
for different temperatures 
at weak disorder $V=1t$ (in (a)) and at strong disorder
$V=4t$ (in (b)). In both cases one finds that $\Lambda^{\rm L}\equiv 
\Lambda_{xx}(q_x\rightarrow 0)=K_x$ at all $T$, verifying the gauge
invariance condition and providing a non-trivial check of our numerics.

\subsection{Transverse response: Superfluid Stiffness}

The transverse response is given by
\begin{equation}
\Lambda^{T}\equiv 
{\rm lim}_{q_y \rightarrow 0} \hskip0.1in   
{\Lambda}_{xx} (q_{x}=0,q_{y},\omega_n=0) \  \cdot
\label{eq:transverse}
\end{equation}
\begin{figure}
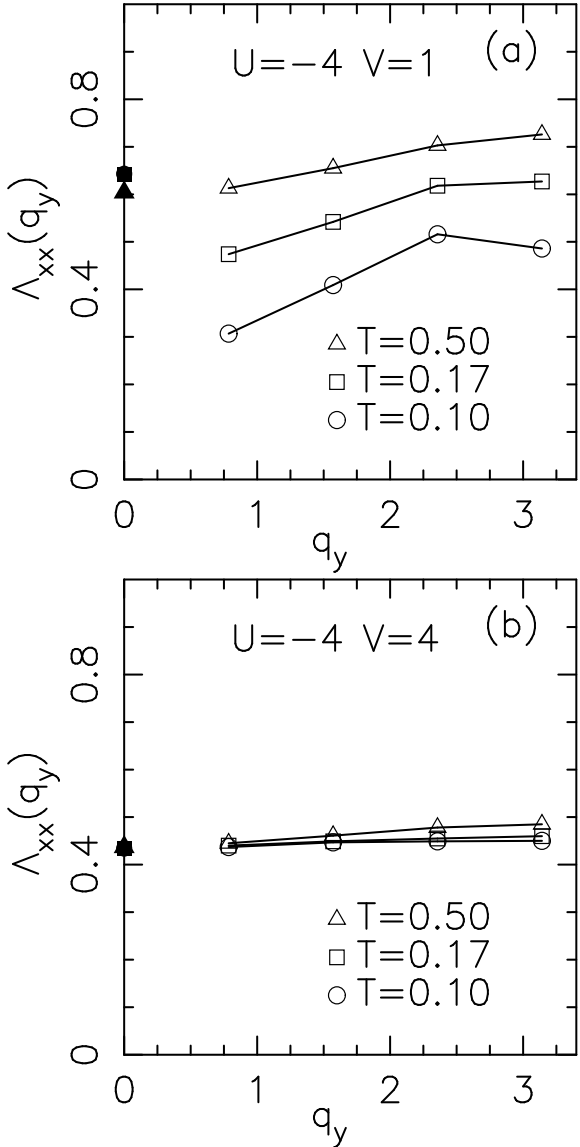

\vskip0mm
\hspace*{0mm}
\psfig{file=lamTV1.ps,height=3.0in,width=3.0in,angle=-90}  
\vskip0mm
\hspace*{0mm}
\psfig{file=lamTV4.ps,height=3.0in,width=3.0in,angle=-90}  
\vskip0mm
\caption{
The transverse current-current correlation function $\Lambda_{xx}(q_y)$
defined 
in Eq.~\ref{eq:transverse}
as a function of $q_{y}$
at $T=0.5t$ (open triangles), $0.17t$ (open squares), and $0.1t$
(open circles).
The corresponding filled points at $q_y=0$ are the 
magnitude of the kinetic energy along $x$ at those temperatures.
For weak disorder $V=1t$ as in (a), 
$\Lambda^{\rm T}=\Lambda_{xx}(q_{y} \rightarrow 0)<K_x$
indicating the development of a finite superfluid stiffness
$D_{s}$ from Eq.~\ref{eq:ds}
with decreasing T.
For strong disorder $V=4t$ as in (b), $D_{s}=0$ at all T.
}
\label{fig:lamtr}
\end{figure} 

In a system with a broken gauge symmetry, the longitudinal and transverse
responses are no longer equal and their difference is precisely
the superfluid stiffness $D_s$ or the related quantity, superfluid density
$\rho_s$, given by
\begin{eqnarray}
\rho_s = D_{s}/\pi & = & [
\Lambda^{\rm L}-
\Lambda^{\rm T}] \nonumber \\
& =  &
[K_{x}-
\Lambda^{\rm T}]\  \cdot
\label{eq:ds}
\end{eqnarray}
It can be seen from Eq.~\ref{eq:ds} that on a lattice the 
superfluid density at $T=0$ is indeed bounded above by the kinetic energy. 
In recent work\cite{NT-bound} we have obtained an
improved upper bound on $D_s$ in a disordered system
in terms of the local kinetic energy
which highlights the dominance of the weak links in determining the superfluid 
stiffness.

In order to extract the superfluid stiffness $D_s$ from Eq.~\ref{eq:ds}
we must extrapolate $\Lambda_{xx}(q_y)$ to $q_y\rightarrow 0$.
Using general symmetry arguments we have
\begin{eqnarray}
j_\alpha(\bf q) &=& \Lambda_{\alpha\beta} (\bf q) A_\beta (\bf q)\nonumber \\
\Lambda_{\alpha\beta} &=& \left(
\delta_{\alpha\beta} - {{q_\alpha q_\beta}\over {q^2}} \right) \Lambda^T(q^2) + 
{{q_\alpha q_\beta} \over {q^2}} \Lambda^L(q^2) 
\end{eqnarray}
so that the linear term in the expansion of $\Lambda^T$ and $\Lambda^L$
is absent and the lowest order term
is quadratic in $q_y$. 
However, the momentum discretization on an $8\times 8$ lattice is
too coarse to see this quadratic behavior.  

\begin{figure}
\vskip0mm
\hspace*{0mm}
\psfig{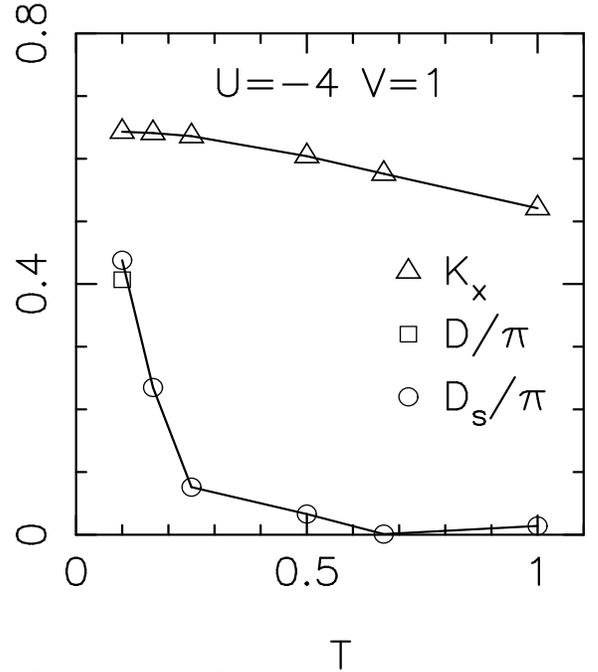}  
\vskip+0mm
\caption{
The superfluid stiffness $D_s$ and $K_x$
as a function of $T$ for $V=1.0$,
$U=-4t$ and $\langle n\rangle=0.875$. 
Also shown is the charge stiffness $D$ at the lowest $T=0.1$.
}
\label{fig:dsT}
\end{figure} 

It is clear from Fig.~\ref{fig:lamtr}
that the transverse correlations behave quite differently 
from the longitudinal correlations.
For weak disorder, at high temperature, $\Lambda^T$ approaches 
$K_{x}$,
but as $T$ is decreased, the two quantities no longer match,
indicating
that a nonzero superfluid
density is developing as shown 
in Fig.~\ref{fig:dsT}.
We see
that $D_{s}$ becomes significantly different from
zero at temperatures $T < 0.2 t$.  This is consistent with
estimates\cite{MOREO-tc} which put $T_c \approx 0.1t$ based on 
a finite size scaling analysis of the pairing correlations,
but seems to contradict recent suggestions
that $T_{c}$ is much lower, approximately $0.03t$.
In Fig.~\ref{fig:dsT} we also show the behavior of $K_x$
which shows no special features as $T$ is lowered.
$K_x$ declines from 0.68 at $V=0$ to 0.39t at $V=5t$, while $D_{s}$
changes by almost two orders of magnitude.
While a reduction in hopping is expected in the presence of 
disorder, the smooth behavior of the kinetic energy emphasizes that
such a local quantity cannot serve as an order parameter for the
localization transition.
When disorder is strong, $\Lambda^T$ remains pinned at $K_{x}$,
for all $T$, suggesting that a superconducting phase is not present.

{\em Thus from the raw data itself there is compelling evidence for a 
superconducting phase at low temperature and at low disorder 
that is qualitatively
distinct from the non-superconducting phase at higher disorder.}

Finally, we note that
the mean field gap is of the order of the hopping integral $t$ for $U=-4t$,
therefore quasiparticle excitations across the gap are suppressed
by a factor $\sim \exp(-t/T)=\exp(-10)$ at a temperature $T=0.1t$.
The finite temperature transition is thus
dominated largely by thermal phase
fluctuations.

\subsection{Superconductor-Insulator Transition}

\begin{figure}
\vskip0mm
\hspace*{0mm}
\psfig{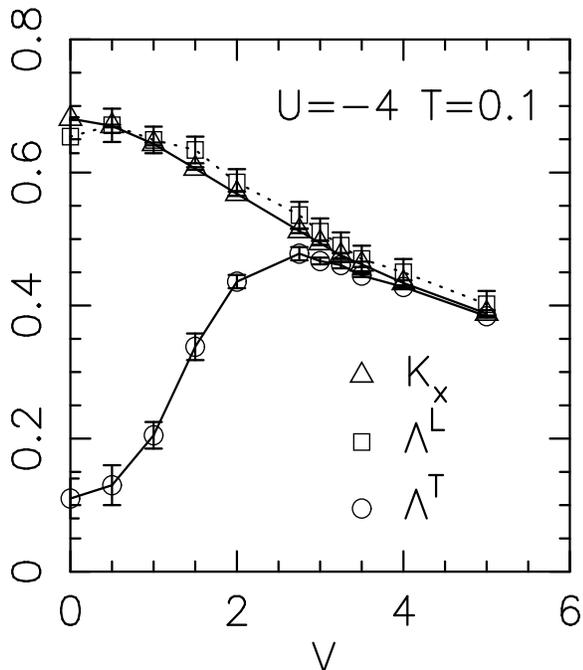}  
\vskip+0mm
\caption{
The transverse current-current correlation function
$\Lambda^T=\lim_{q_y\rightarrow 0}\Lambda_{xx}(q_{y})$
as a function of disorder at $T=0.10$.
Also shown is $K_x=\Lambda^L$, the longitudinal response function,
 as a function of disorder.
The difference between $\Lambda^L$ and $\Lambda^T$
is the superfluid stiffness as seen from Eq.~\ref{eq:ds}.
}
\label{fig:lamtr_V}
\end{figure}

\begin{figure}
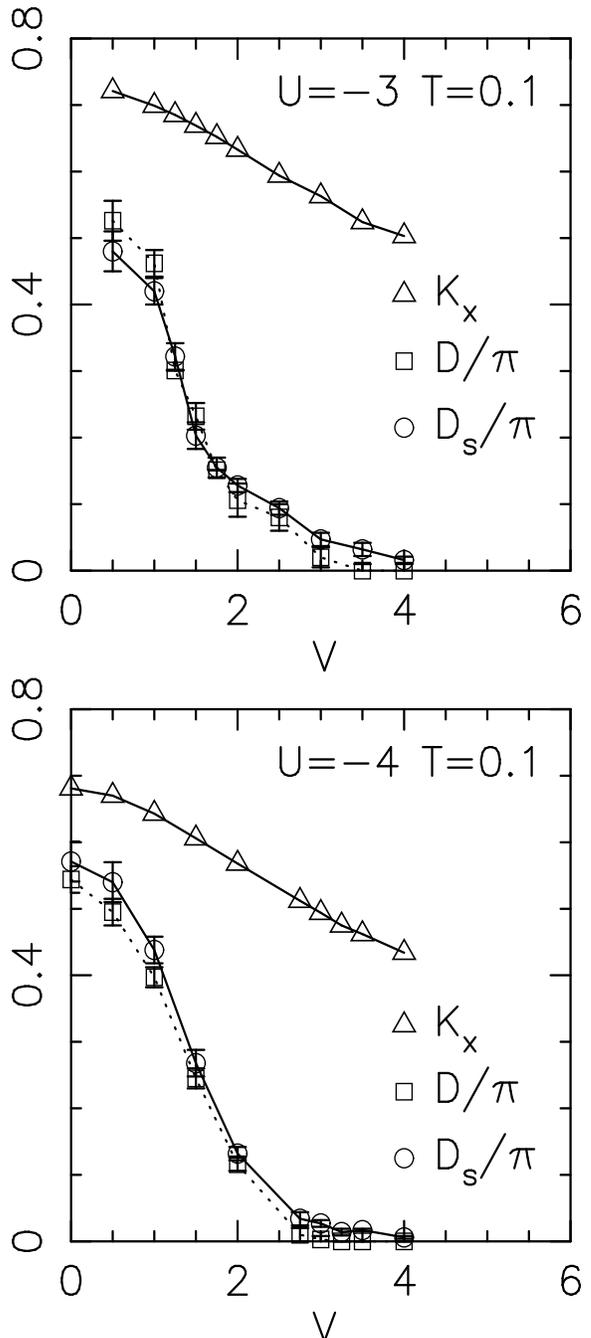

\vskip0mm
\hspace*{0mm}
\psfig{file=DDsvsVU3.ps,height=3.5in,width=3.0in,angle=-90}  
\vskip0mm
\hspace*{0mm}
\psfig{file=DDsvsVU4.ps,height=3.5in,width=3.0in,angle=-90}  
\caption{
The superfluid stiffness $D_{s}$ and the charge stiffness $D$ 
as a function of disorder strength $V$ for $U=-3t$
and $U=-4t$.
Note the rapid suppression with disorder and the transition from a
superconductor to an insulator beyond a critical disorder.
}
\label{fig:dsV}
\end{figure}

In order to determine the location of the transition, we now present data
at a set of disorder values which sweeps through the values $V=1$ -- $4$
which we argued in the preceding section
brackets the transition.
In Fig.~\ref{fig:lamtr_V} we show the 
extrapolated values of $\Lambda_{xx}(q_y)$ and $K_x$ as a function of disorder.
It is evident that the transition is driven by 
the variation of $\Lambda^T$.
In Fig.~\ref{fig:dsV} we show $D_s$ as a function of disorder strength 
at fixed temperature $T=0.1t$, for $U=-3t$ and $U=-4t$.  
The decrease in $D_s$ with increasing disorder is 
consistent with the decline in the order parameter shown in
Fig.~\ref{fig:pairmax}.

The superfluid stiffness
$D_{s}\sim \delta^\zeta$ where $\delta=|V-V_c|/|V_c|$ is the distance 
from critical disorder. The exponent $\zeta$ is expected to be larger than 
unity since $\zeta =z \nu$ and in 2d it has been argued that
$\nu\ge 2/d=1$
and $z=2$.
A value of $\zeta>1$ implies that the finite size rounding will shift the
critical point on the infinite lattice to {\it higher} values compared to the
point where $D_s$ becomes small on finite lattices. So we expect 
that the critical point for the SIT
may lie around $V_{c} \approx$ 3--4t for $U=-4t$.

\begin{figure}
\vskip0mm
\hspace*{0mm}
\psfig{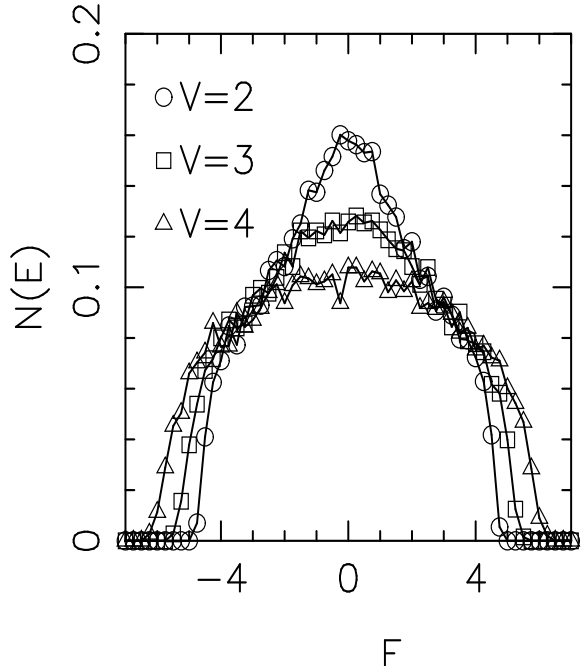}  
\caption{
The density of states $N(E)$ 
as a function of energy $E$
for noninteracting fermions ($U=0$) on an $8\times 8$ lattice at
density $\langle n\rangle=0.875$ for disorder strengths around
$V=2t,\ 3t<V_c$ and $V=4t>V_c$, where $V_c$ is the critical disorder of 
the interacting problem with $U=-4t$.
}
\label{fig:dos}
\end{figure} 

\begin{figure}
\vskip0mm
\hspace*{0mm}
\psfig{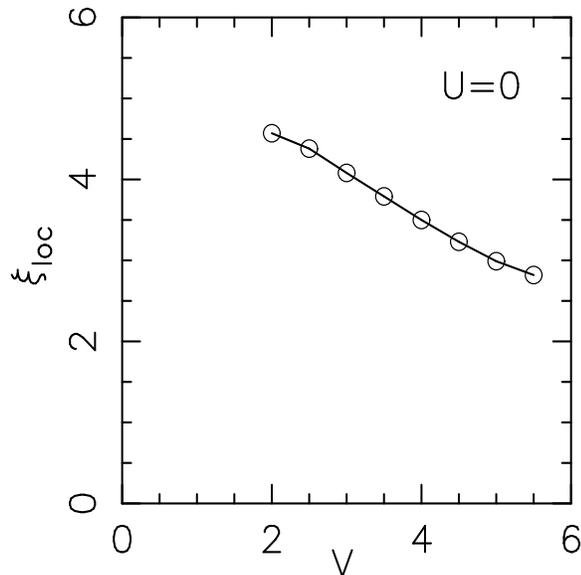}  
\caption{The approximate localization length of the eigenstate at the
Fermi surface inferred from the participation ratio by
as a function of disorder strength
$V$.
We see that the single
particle eigenstates do not show any sharp behavior
around the critical disorder $V_{c}\sim 3.25$ found 
for the SIT in our
QMC simulations of the interacting problem.} 
\label{fig:xiloc}
\end{figure} 

It is reasonable to ask to what extent the sharp drop in the pair 
correlations and the transition to insulating behavior
in the resistivity might reflect changes in the noninteracting eigenstates of
the Hamiltonian.  Is the fact that the pairing correlations are robust 
at $V=0$ but zero at $V=5t$ a consequence of some 
changes in the extent of the single-particle wavefunctions due to disorder?

In Fig.~\ref{fig:dos} we show the density of states $N(E)$ 
for $U=0$ and different amounts of randomness bracketing
$V_c$.  We see that disorder broadens $N(E)$, as expected,
but the behavior of this
quantity through $V_c$ is smooth.  

We show in Fig.~\ref{fig:xiloc} the
localization length or the
``size'' of the eigenstate at the Fermi surface, defined by
$\xi_{loc} = \sqrt{{\cal PR}(E_{F})}$ as a function of disorder strength.
$\xi_{loc}$ shows a smooth  decrease as a function of $V$, without any sharp
feature at $V_{c}$.
We conclude that the SIT
is not occuring as a consequence of a $U=0$ 
Anderson transition on the finite lattice,
even though the wavefunctions are localized on the scale of the
linear lattice size $L$.  
Instead, the transition is a
genuinely nontrivial many body effect.

\subsection{Coherence length}

In principle, we can extract the superconducting
coherence length $\xi$ for the many body problem
from the 
dependence of $\Lambda^T(q_{y})=a + b q_y^2$ 
for small $q_y$. From Eq.~\ref{eq:ds} we see that
\begin{equation}
{{D_s(q_y)}\over {\pi}}={{D_s}\over {\pi}}\left[1-q_y^2\xi^2\right]
\label{eq:xi}
\end{equation}
where $\xi^2=b/(D_s/\pi)$. As a function of disorder $\xi$ is found to 
decrease slightly for low disorder and is expected to diverge as the critical
disorder is approached. However, it is difficult to deduce such a divergence 
from the data since both $b$ and $D_s$ are becoming small near the transition.
Further work on this problem is required, since it would be useful to
obtain the coherence length to track the quantum phase transition.

\section{Charge Stiffness}

A superconductor is characterized by the Meissner effect, measured by the 
superfluid stiffness, as well as by an infinite conductivity. 
A signature of the latter 
is a delta function in the optical conductivity
\begin{equation}
{\rm Re}\sigma(\omega) \ = \ D \delta(\omega) + {\rm Re} \sigma_{reg}(\omega)
\label{eq:sigma1}
\end{equation}
with weight 
$D = \pi [ K_{x} - {\rm lim}_{\omega \rightarrow 0} \hskip0.02in
{\rm Re} \hskip0.02in
{\Lambda}_{xx} ({\bf q}=0;\omega + i0^{+}) ]$, known as the charge stiffness.
The regular part of the conductivity is given by
(suppressing the ${\bf q}=0$ and omitting the $xx$ subscripts)
\begin{equation}
{\rm Re} \sigma_{\rm reg}(\omega) = { {{\rm Im}\Lambda(\omega)}\over {\omega}}
\label{eq:sigma2}
\end{equation}
where
$\Lambda(\omega + i0^+)  = {\rm Re}\Lambda(\omega)
+ i\ImL(\omega)$.

In order to obtain the dc limit we proceed as follows. 
We start with the sum rule
\begin{equation}
\int_0^\infty d\omega {\rm Re}\sigma(\omega) = {\pi\over 2} K_x
\label{eq:sumrule}
\end{equation}
and combine with Eq.~\ref{eq:sigma1} to get
\begin{equation}
\int_0^\infty d\omega {\rm Re}\sigma_{reg}(\omega) = 
{\pi\over 2} K_x-{D\over 2}.
\label{eq:sumrule2}
\end{equation}
Next, using the spectral representation for $\Lambda(z)$,
\begin{equation}
\Lambda(z) = \int_{-\infty}^{\infty} {{d\omega}\over \pi}
{{\rm Im \Lambda(\omega)}\over{\omega -z}}\  
\label{eq:spectral}
\end{equation}
and substituting $z=i\omega_n$ we get,
\begin{equation}
\Lambda(\omega_n) = {2\over \pi}\int_{0}^{\infty} {d\omega}
{{\omega \rm Im \Lambda(\omega)}\over{\omega^2 +\omega_n^2}}\  \cdot
\label{eq:lambda_om}
\end{equation}
Using Eq.~\ref{eq:sigma2}, 
\begin{eqnarray}
\Lambda(\omega_n) &=&  
{2\over \pi} \int_0^\infty d\omega {\rm Re} \sigma_{\rm reg}(\omega)
\\
&-& {2\over \pi} \omega_n^2\int_0^\infty d\omega 
{{{\rm Re} \sigma_{\rm reg}(\omega)}
\over{\omega^2 + \omega_n^2}}\  \cdot
\label{eq:lambda_om2}
\end{eqnarray}
Substituting for the first term from Eq.~\ref{eq:sumrule2} and defining 
the Matsubara correlation function
\begin{equation}\label{eq:dom}
D(\omega_n) = \pi[K_x - \Lambda(\omega_{n})]
\end{equation}
whence,
\begin{equation}\label{eq:dom2}
D(\omega_n) = D + 2\omega_n^2\int_0^\infty d\omega {{\sigma_{\rm reg}(\omega)}
\over{[\omega^2 + \omega_n^2]}}\  \cdot
\end{equation}

\begin{figure}
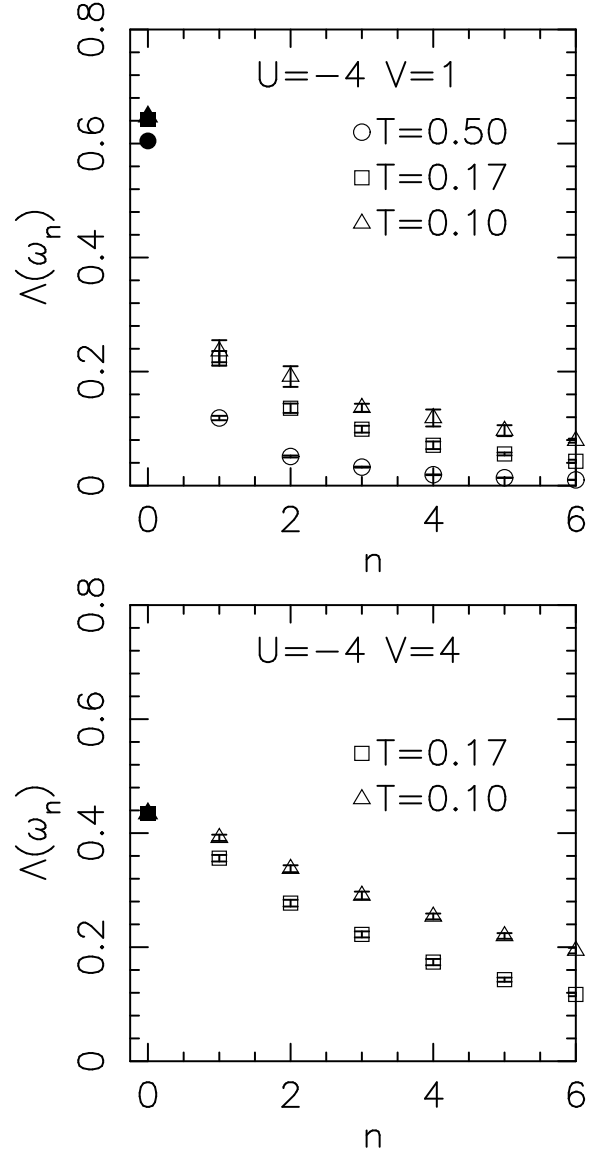

\vskip-00mm
\hspace*{0mm}
\psfig{file=lamom_v1.ps,height=3.0in,width=3.0in,angle=-90}  
\vskip-0mm
\hspace*{0mm}
\psfig{file=lamom_v4.ps,height=3.0in,width=3.0in,angle=-90}  
\caption{
The behavior of 
$\Lambda(\omega_n)$
as a function of $n$
at temperatures $T=$ 0.17 and 0.10.
The corresponding filled points at $n=0$ are the 
values of $K_x$
at those temperatures.
For weak disorder $V=t$ 
$\lim_{\omega_n\rightarrow 0} \Lambda(\omega_n)< K_x$
indicating a finite charge stiffness $D$ whereas at strong
disorder $V=4t$,
$\lim_{\omega_n\rightarrow 0} \Lambda(\omega_n)\approx K_x$
implying that $D=0$.
}
\label{fig:lamom}
\end{figure}

\begin{figure}
\vskip-20mm
\hspace*{-10mm}
\psfig{file=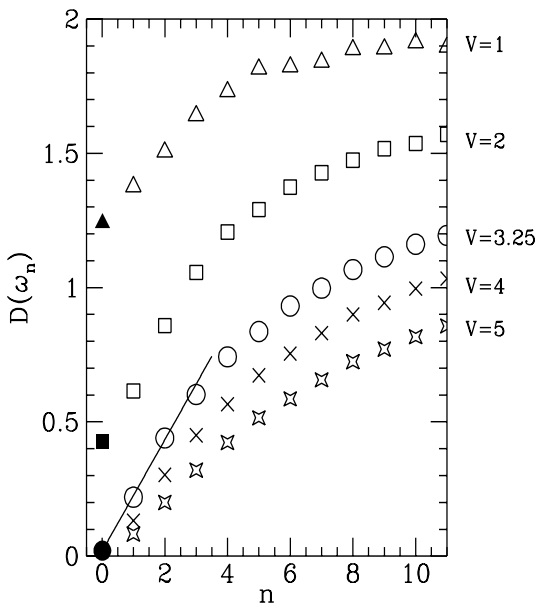,height=4.5in,width=4.0in,angle=-0}  
\caption{
The behavior of $D(\omega_n)$ defined in Eq.~(\ref{eq:dom})
as a function of $n$ for $T=0.1$, $U=-4t$ and $\langle n \rangle=0.875$
for disorder strengths $V=1,2,3.25,4,5t$.
The corresponding filled points at $\omega_n=0$ are the 
extrapolated values of the charge stiffness $D$
at those $V$. The critical disorder $V_c=3.25$ is identified by the vanishing
of $D(\omega_n)$. The straight line is a linear fit to the low $\omega_n$ data
whose slope is proportional to the critical conductivity at the transition
from Eq.\ref{eq:cond2}}.
\label{fig:dom}
\end{figure}

The behavior of $\Lambda(\omega_n)$ as a function of 
$n$ is shown in
Fig.~\ref{fig:lamom} for low disorder $V=1t$ in (a) and
for high disorder $V=4t$ in (b).
The behavior of $\Lambda(\omega_n)$ is qualitatively similar to
Fig.~\ref{fig:lamtr}.  
That is, at strong disorder, $\Lambda(\omega_n\rightarrow 0)\approx K_x$ 
at all temperatures and according to Eq.~\ref{eq:dom} 
this implies the charge stiffness $D\approx 0$,
as is the superfluid stiffness $D_s$.
At weak disorder and at low T
on the other hand, $\Lambda(\omega_n\rightarrow 0) < K_x$ 
implying that $D$ is nonzero.
In Fig.~\ref{fig:dom} we show
$D(\omega_n)$ as a function of $n$ which is found to
increase monotonically with $n$
from $D(\omega_n \rightarrow 0) = D$ to
$D(\omega_n \rightarrow \infty) = \pi(-K_x)$ (not shown in the figure)
but verified in the data.

The behavior of $D$ as a function of disorder extracted from Eq.~\ref{eq:dom2}
is shown in Fig.~\ref{fig:dsV}. 
We see that $D$ and $D_s$ are within
10-20\% of each other for all the parameters shown.
Thus there is remarkable consistency between the
superfluid stiffness $D_s$ and the strength $D$
of the delta function in the optical conductivity, obtained
from two very different correlation functions.

Do these techniques give sensible results in the noninteracting, clean
limit?  For $U=V=0$ we find
the charge stiffness $D/\pi= 0.79=\langle -K_x\rangle$ 
whereas the superfluid stiffness
$D_{s}/\pi=0.0243$ for filling $\langle n \rangle=0.86$ and $T=0.1t$ 
on an $8\times 8$ system.
Thus our numerics are correctly telling us that free fermions are metallic
with a nonzero $D$, but a very small
$D_s$, which will go to zero as the system size increases.

While the approximate 
equality of $D$ and $D_s$ in Fig.~\ref{fig:dsV} for a superconductor
is a good check on the calculation, it
emphasizes that the charge stiffness D at $T=0$ cannot be used to
characterize the non-superconducting state for $V\geq V_c$ since 
neither dirty metals nor insulators have a $\delta$ function in 
$\sigma(\omega)$ at $\omega=0$.
Hence we turn to the conductivity.

\section{Conductivity}

\smallskip

The dc conductivity $\sigma_{dc}={\rm lim}_{\omega \rightarrow 0} {\rm Re}
\sigma_{reg}(\omega)$ defined in Eq.~\ref{eq:sigma2} 
is of considerable theoretical and experimental interest 
as it distinguishes the two non--superconducting
phases-- metal (above $T_c$) vs insulator.  
The fluctuation-dissipation theorem
relates ${\rm Im}\Lambda(\omega)$ which is required 
for the calculation of $\sigma_{dc}$,
to $\Lambda(\tau)$ which is obtained from QMC data by
\begin{equation}
\Lambda(\tau) = \int^{+\infty}_{-\infty} {d\omega \over \pi}
{{\exp(-\omega\tau)}\over{\left[1 - \exp(-\beta\omega)\right]}}
\ImL(\omega)
\label {eq:fd}
\end{equation}
valid for $0\le\tau\le\beta$. 
However, the evaluation of ${\rm Im}\Lambda(\omega)$ requires an
analytic continuation of noisy imaginary time data\cite{maxent} 
which is
difficult.
We derive below an 
approximate expression for $\sigma_{dc}$\cite{NT-sit},
analogous to that introduced previously
for the susceptibility\cite{NT-pseudogap1},
by noting that if one sets $\tau=\beta/2$, the kernel in
Eq.~\ref{eq:fd} cuts off contributions from high frequencies, and 
the important range of $\omega$ is restricted to increasingly small
values as $\beta$ becomes large.
Therefore, at low enough temperatures
one might replace
\begin{figure}
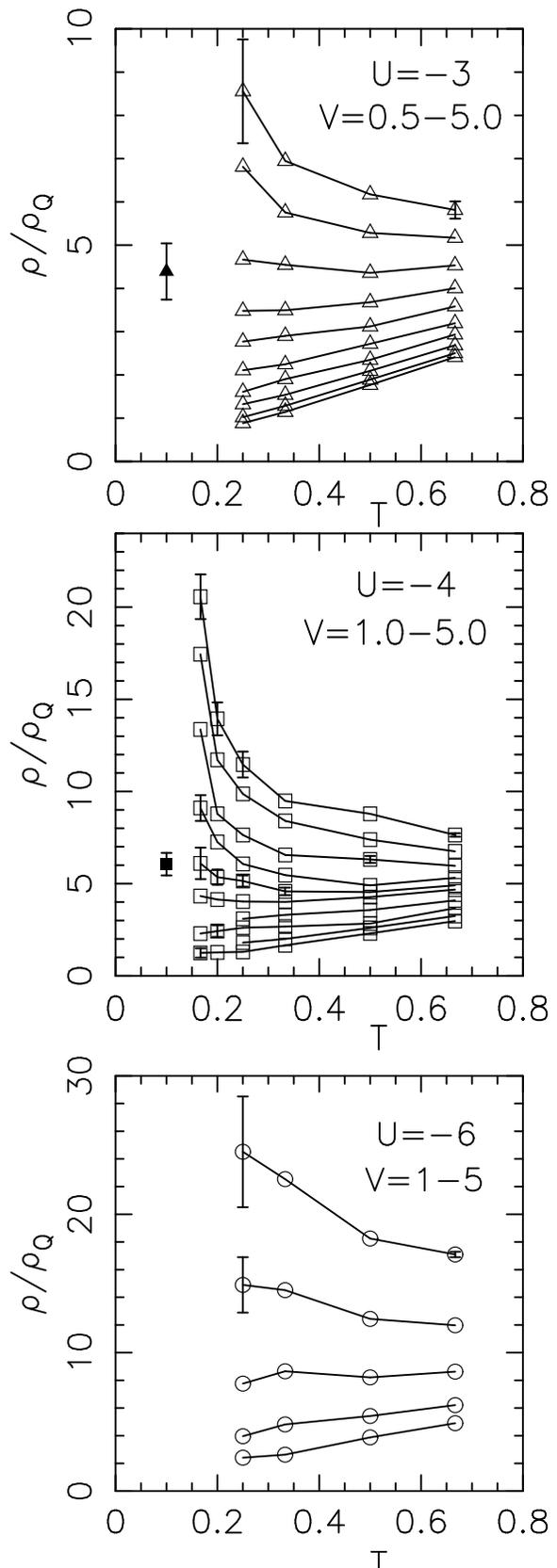

\vskip0mm
\hspace*{0mm}
\psfig{file=rhoTU3.ps,height=2.9in,width=3.0in,angle=-90}  
\vskip0mm
\hspace*{0mm}
\psfig{file=rhoTU4.ps,height=2.9in,width=3.0in,angle=-90}  
\vskip0mm
\hspace*{0mm}
\psfig{file=rhoTU6.ps,height=2.9in,width=3.0in,angle=-90}  
\vskip0mm
\caption{
The behavior of the resistivity $\rho$ obtained from Eq.~\ref{eq:ac} 
as a function of $T$ for various disorder strengths. Figs. 
correspond to $U=-3,-4,-6$ respectively.
}
\label{fig:rhoT}
\end{figure} 
\noindent $\ImL(\omega) \simeq  \omega\dc$ over the entire range of integration,
which leads to the result 
\begin{equation}
\dc = {{\beta^2\Lambda(\tau=\beta/2)}\over{\pi}}\  \cdot
\label {eq:ac}
\end{equation}
Note that Eq.~\ref{eq:ac}
is only valid in the normal state ($T_c\approx 0.1t$)
where ${\rm Im} \Lambda(\omega)\sim \omega\sigma_{\rm dc}$ at low
frequencies.
We will present a number of self-consistent checks of
Eq.~\ref{eq:ac} in the metallic state above $T_c$ 
of the superconductor and the localized phase. We defer a discussion 
of the extraction of the conductivity at the SIT to the next section.

If Fig.~\ref{fig:rhoT} we show the behavior of the  
resistivity $\rho=1/ \sigma_{\rm dc}$
obtained from Eq.~\ref{eq:ac} 
as a function of temperature. The resistivity
shows a behavior qualitatively similar to that seen in experiment:  
when the control parameter, in this case disorder, 
is weak, the behavior is metallic and $\rho$ decreases
as $T$ decreases. On the other hand, 
for strong disorder, the behavior is insulating and $\rho$
increases as $T$ decreases.
Our plots are qualitatively
similar to those observed experimentally, though the experimental range of
resistivities is much greater.

As is often done experimentally, data for $\rho(T)$ at different
$V$ can be replotted to show $\rho(V)$ for different 
temperatures.
For $V<V_{c}$ the resistivity decreases as $T$ is lowered, while
for $V>V_{c}$ the resistivity increases as $T$ is lowered.  This leads
to a characteristic crossing pattern in $\rho(V)$ which
allows for an estimation of the critical amount of disorder $V_{c}$
as well as the critical resistance $\rho(V_{c})$ 
at the transition.
Note that the crossing pattern does not follow from any
deep scaling principle. 
Instead, it is merely a consequence of the monotonicity of the
plots of $\rho(T)$ for a given $V$, which, to within error bars, either
steadily increase or decrease as $T$ is changed.

\begin{figure}
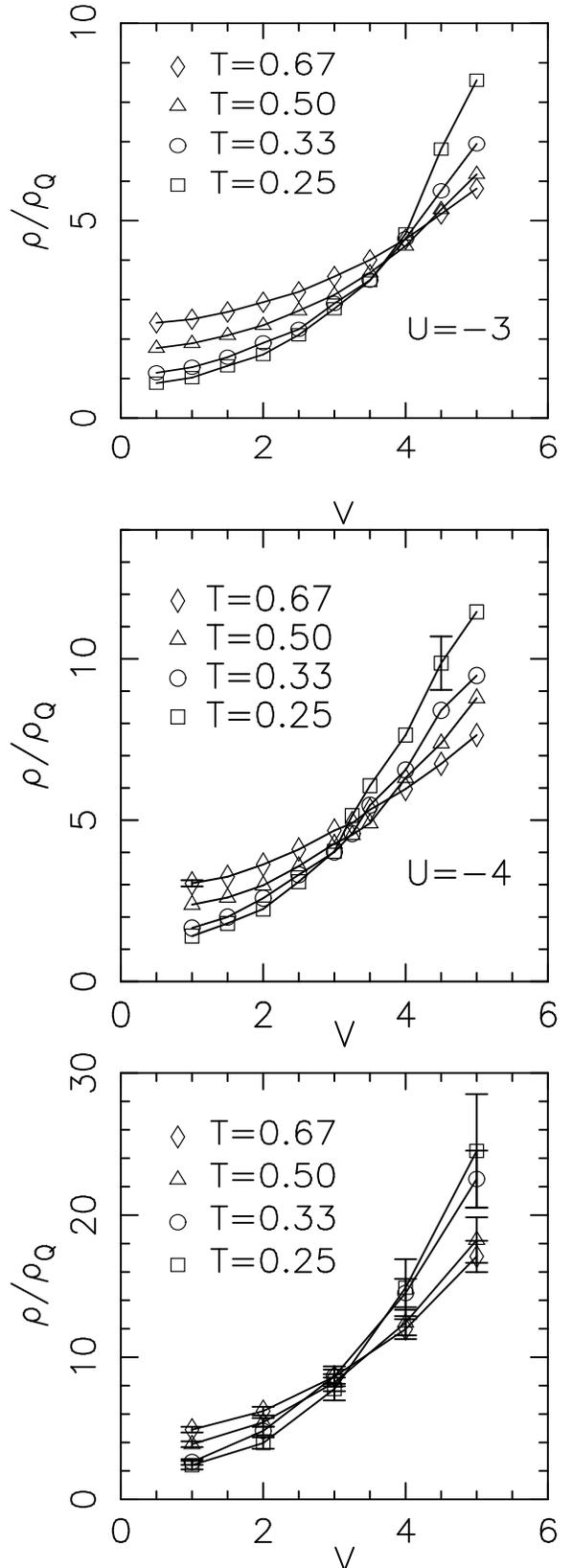

\vskip0mm
\hspace*{0mm}
\psfig{file=rhoVU3.ps,height=2.85in,width=3.0in,angle=-90}  
\vskip0mm
\hspace*{0mm}
\psfig{file=rhoVU4.ps,height=2.85in,width=3.0in,angle=-90}  
\vskip0mm
\hspace*{0mm}
\psfig{file=rhoVU6.ps,height=2.85in,width=3.0in,angle=-90}  
\vskip0mm
\caption{
The behavior of the resistivity $\rho$ obtained from Eq.~\ref{eq:ac} 
as a function of $V$ for various temperatures.
Figs. correspond to $U=-3,-4,-6$ respectively.
}
\label{fig:rhoV}
\end{figure}

\begin{figure}
\vskip-00mm
\hspace*{-00mm}
\psfig{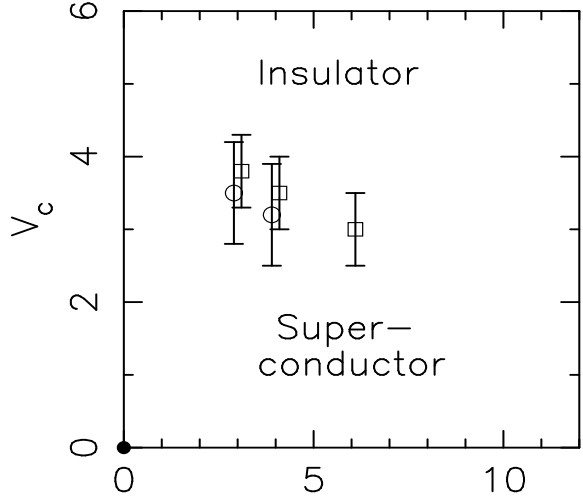}  
\vskip-00mm
\caption{
The locus of critical disorder $V_c(U)$ 
for intermediate couplings in the disorder $V$ -- attraction $|U|$ plane.
The $V_c$ values are obtained by two independent methods,
the vanishing of superfluid
density $D_s$ (open circles), and the crossing of the resistivity 
$\rho$ (open squares).  The filled circle at $U=V=0$ emphasizes
that all noninteracting 
states are localized for any non--zero disorder $V$ in
two dimensions.}
\label{fig:pd2d_mc}
\end{figure}

From Fig.~\ref{fig:dsV} and Fig.~\ref{fig:rhoV}
we see clear evidence for a SIT
at a critical disorder $V_c(U)$ whose dependence 
on the strength of the attraction
is shown in Fig.~\ref{fig:pd2d_mc}.

It has recently been emphasized by Sachdev\cite{SACHDEV-pvt}
that using Eq.~\ref{eq:ac} to extract the resistivity is not applicable near
a quantum phase transition as there is {\em no scale in the problem}.
Note that it was assumed in the derivation of Eq.~\ref{eq:ac} that below some 
scale which was independent of $T$, it was possible to assume that
${\rm Im} \Lambda(\omega)\sim \omega\sigma_{\rm dc}$. This assumption
breaks down near a quantum critical point since by definition all scales
become soft.  Away from the transition, 
Eq.~\ref{eq:ac} gives a good description of $\rho_{dc}(T)$, however,
close to the transition, it cannot be used to extract
the critical conductivity. The agreement of the transition point
obtained by the conductivity crossing plots and the
measurements of the superfluid and charge stiffness 
suggest that Eq.~\ref{eq:ac} has a useful range of validity.

We discuss another potential method to extract the conductivity at the critical
point.
As seen in Fig.~\ref{fig:dom} at a critical disorder $D$ vanishes. 
At this disorder
assume that ${\rm Re} \sigma(\omega) \rightarrow \sigma_0 = const $,
for frequencies $\omega < \omega_c$, a cut-off value. 

Then from Eq.~\ref{eq:dom2}
\begin{eqnarray}
D(\omega_n) &=&  2\omega_n^2\int_0^\infty d\omega {{\sigma_{\rm reg}(\omega)}
\over{\omega^2 + \omega_n^2}}\nonumber \\
	&=& 2 \sigma_0 |\omega_n| \tan^{-1}\left({{\omega_c}\over{\omega_n}}
\right) + 
 2\omega_n^2\int_{\omega_c}^\infty d\omega {{{\rm Re}\sigma(\omega)}
\over{\omega^2 + \omega_n^2}}
\label{cond1}
\end{eqnarray}
which in the limit of small Matsubara frequencies is given by
\begin{equation}
D(\omega_n) = \pi \sigma_0 |\omega_n| + {\cal O}(\omega_n)^2\  \cdot
\label{eq:cond2}
\end{equation}

\noindent
The conductivity at the critical point obtained from
Eq.~\ref{eq:cond2} and from the crossing of the resistivity curves
described above are in agreement to about 10 \%.

\begin{figure}
\vskip-00mm
\hspace*{-0mm}
\psfig{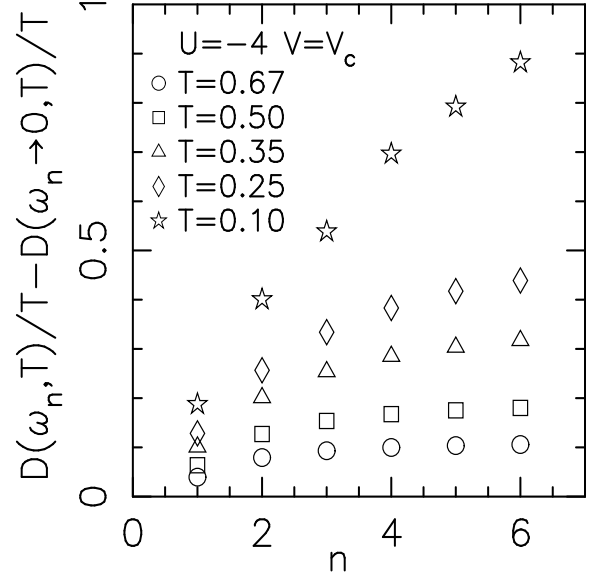} 
\vskip04mm
\caption{
The behavior of $F(n)=D(\omega_n,T)/T$ $ - D(\omega_n\rightarrow 0,T)/T$ 
as a function of $n=\omega_n/(2\pi T)$ 
defined in Eq.~(\ref{eq:noscale}) at the critical disorder $V_c\sim 3.25t$
and various $T$.}
\label{fig:noscale}
\end{figure}

\begin{figure}
\vskip-00mm
\hspace*{-0mm}
\vskip04mm
\psfig{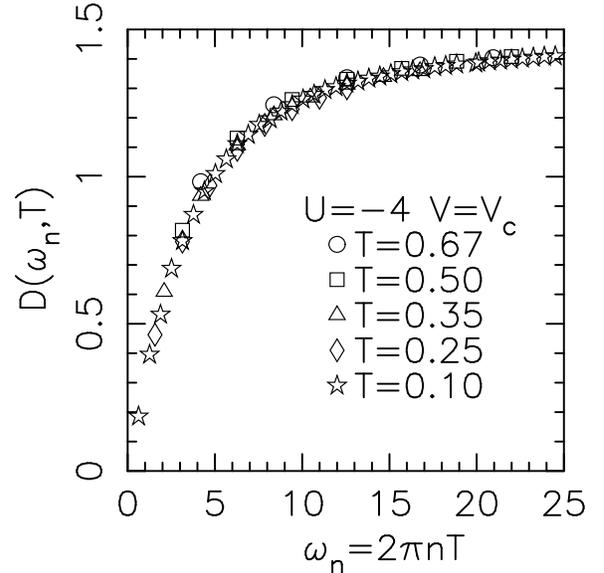}  
\caption{
The behavior of $D(\omega_n,T)$ 
as a function of $\omega_n$ 
at the critical disorder $V_c\sim 3.25t$
and various $T$.}
\label{fig:scale}
\end{figure}

Near a quantum critical point, we expect 
\begin{equation}
\sigma_{reg} (\omega,T,V=V_c) = \sigma_Q \sigma(\omega/T)\  \cdot
\label{eq:sign}
\end{equation}
\noindent From Eq.~\ref{eq:dom2}, this implies that
\begin{eqnarray}
{{D(\omega_n)}\over{T}} &=& {{D(T)}\over T} + 8 \pi^2 n^2\sigma_Q\int_{x_c}
^\infty dx 
{{f(x)}\over{x^2 + 4 \pi^2 n^2}} \nonumber \\
&\equiv & G(T) + F(n)
\label{eq:noscale}
\end{eqnarray}
where $x=\omega/T$. Thus $D(\omega_n)/T$ is a sum of two terms-- the first one
$G(T)=D(T)/T$ is only a function of $T$ and the second term $F(n)$
is only a function of $n$, with $F(n\rightarrow 0) = 0$.
We set $V=V_c\sim 3.25t$ and by extrapolating the behavior of $D(\omega_n)/T$
to $n\rightarrow 0$ obtain $G(T)$. 
In Fig.~\ref{fig:noscale}, we show the behavior of $F(n)$ vs $n=\omega_n/
2 \pi T$ at the critical point for various temperatures.
The data are not found to scale, unlike our expectations at a critical
point. Instead if we plot $D(\omega_n)$ vs $\omega_n$ we see a 
remarkable scaling behavior of the data for various temperatures
as seen in Fig.~\ref{fig:scale}.
It is not really clear as to why the data when plotted as in 
Fig.~\ref{fig:noscale} does not scale.


It has been claimed in Ref.~\cite{DAMLE-SACHDEV} that 
since in QMC the lowest frequency that can be accessed is
$\omega_1=2\pi T > T$, it is not possible to extract the dc resistivity 
using Eq.~\ref{eq:cond2} in
the low frequency limit.
While this objection appears very sound,
it is nevertheless the case that the conductivity inferred 
from Eq.~\ref{eq:cond2}
including its values in the vicinity of the critical point,
is consistent with many other, completely rigorously founded, aspects of
our simulation.  By this we mean that the location of the transition
inferred from the analysis of the data using Eq.~\ref{eq:cond2}
is in remarkable
agreement with the location obtained from the superfluid stiffness
$D_s$, and the charge stiffness $D$. 
Furthermore, the value of the conductivity at the transition 
is consistent with the value obtained from
Eq.~\ref{eq:ac}.
At present we do not understand fully why the method appears to
be so consistent with our other data despite the objections raised in
Ref.~\cite{DAMLE-SACHDEV}.
\section{Conclusions}

We have studied the effect of disorder on an s-wave superconductor
of fixed coupling strength (modeled as an attractive Hubbard model
away from half filling). We have found that with increasing disorder,
the superfluid stiffness (obtained from the transverse current-current
correlation function) and the charge stiffness (obtained from the 
$\tau-$ dependent current-current correlation function, vanish at a 
critical disorder, signaling a transition to a localized phase.
The importance of our work lies in the fact that the SIT which has been
observed experimentally, has eluded all mean field treatments of the
problem. Ours is the first theoretical study of a fermionic model
to obtain a {\em transition}
between the superconductor and localized phases upon increasing the 
disorder strength.

\section{Outstanding questions}

Having established the existence of the SIT the outstanding questions
now relate to obtaining a quantitative characterization of the transition. 
For this it is necessary to perform finite
size scaling in both the spatial $(L\rightarrow \infty)$ and the temporal
$(T\rightarrow 0)$ dimensions to obtain the location of the critical disorder
from the vanishing of the superfluid stiffness $D_s$ as well as the 
vanishing of the charge stiffness $D$. 
From the scaling of the data it is then possible to extract the 
dynamical exponent $z$ and the correlation length exponent $\nu$.
Such an analysis will tell us
whether the fermion SIT is in the 
same universality class as the bosonic superfluid-insulator transition or not.
While there have been several studies of the 
SF-I transition\cite{FISHER-bosons,ZIMANYI-bosons} in the 
boson Hubbard model\cite{RTS-bosons,NT-bosons,RUNGE} and
its variants\cite{SORENSEN}, we believe that the situation with regard to 
the value of the 
exponents is still unclear\cite{NT-issues}. 
This is largely because of the complications of
finite size scaling analysis inherent in a quantum phase transition that 
necessarily involves two variables--(system size
$L\rightarrow \infty$ and temperature $T\rightarrow 0$).

Once the location and exponents characterizing the transition are
determined, the key question is the value of 
the resistivity at the transition and the possibility
of its universality.
There is some experimental 
evidence that despite the
wide range of materials and control parameters, the 
value of the resistance right at the transition $R^\ast$ is 
always quite close to the ``universal'' 
value\cite{GOLDMAN,DYNES,HEBARD,VALLES} $R_{Q} = h
/ 4e^{2}$. While there is still some debate
concerning whether this number is truly the same
for all systems, it is certainly clear that the variation
in $R^\ast$ is much less than the variations 
in the location of the transition in other control parameters such as
the temperature, magnetic field strength, or film thickness.
Recent experiments of Yazdani {\it et al.}\cite{YAZDANI}
have interpreted the variation in $R^\ast$ that exists
in terms of separate bosonic and fermionic contributions to the
resistivity.  Thus, calculations with models which include
electronic degrees of freedom like the attractive Hubbard Hamiltonian
are needed to supplement work on bosonic theories.
To address this set of issues concerning $R^{\ast}$,
we require an exact method to calculate the 
resistivity at the transition, as would be provided by
maximum entropy techniques.
We are currently working on this problem.


\smallskip

\noindent
\underbar{Acknowledgements:}
We would like to thank M. Randeria and S. Sachdev for many useful
discussions.
We also thank K.~Runge for providing useful scripts for
doing the disorder averaging.
The numerical calculations were performed 
at the NCSA.
This work was supported by the
NSF under grant No.~DMR95-28535 (R.T.S.).


\begin{references}
\bibitem{PH} This is seen by performing a phased particle-hole transformation
on the down spin electrons which results in a mapping onto the
repulsive model.
\end{references}
\end{document}